\documentclass{article}
\usepackage{spconf,amsmath,graphicx,hyperref}
\usepackage{booktabs} 
\usepackage{caption}
\usepackage{subcaption}

\usepackage[some,bottom]{background}

\SetBgContents{
\begin{minipage}{10in}
\begin{center}
{K. Choi and M. Kim, ``A Comparative Analysis of Poetry Reading Audio: Singing, Narrating, or Somewhere in Between?"}\\ 
{IEEE International Conference on Acoustics, Speech and Signal Processing (ICASSP), Seoul, Korea, 2024, pp. 1296-1300, doi: 10.1109/ICASSP48485.2024.10447582.}        
\end{center}
\end{minipage}
}
\SetBgScale{.8}
\SetBgOpacity{1}
\SetBgColor{black}
\SetBgVshift{1cm}

\newcommand{\argmax}{\operatornamewithlimits{\arg \max}}
\title{A Comparative Analysis of Poetry Reading Audio:\\Singing, Narrating, or Somewhere In Between?}
\name{Kahyun Choi$^1$\thanks{This work was supported by RE-252382-OLS-22 from the Institute of Museum and Library Services.}\thanks{The authors appreciate Smule, Inc. for providing the Intonation dataset and Sumitha Vellinalur Thattai's help on the initial data collection effort.} and Minje Kim$^2$\sthanks{Work done at Indiana University.}}
\address{$^1$Indiana University, Luddy School of Informatics, Computing and Engineering, Bloomington, IN, 47408\\
$^2$University of Illinois at Urbana-Champaign, Department of Computer Science, IL, USA 61801}
\begin{document}
\ninept
\maketitle
\begin{abstract}
This paper provides a computational analysis of poetry reading audio signals at a large scale to unveil the musicality within professionally-read poems. Although the acoustic characteristics of other types of spoken language have been extensively studied, most of the literature is limited to narrative speech or singing voice, discussing how different they are from each other. In this work, we develop signal processing methods, which are tailored to capture the unique acoustic characteristics of poetry reading based on their silence patterns, temporal variations of local pitch, and beat stability. Our large-scale statistical analyses on three big corpora, each of which consists of narration (LibriSpeech), singing voice (Intonation), and poetry reading (from The Poetry Foundation), discover that poetry reading does share some musical characteristics with singing voice, although it may also resemble narrative speech.
\end{abstract}
\begin{keywords}
Poetry reading, singing voice, pitch tracking, beat tracking, musicality
\end{keywords}



\BgThispage
\section{Introduction}
\label{sec:intro}

Recently, poetry reading has gained increased popularity. According to the 2017 survey from the National Endowment for the Arts, the poetry readership in the United States increased by 76\% between 2012 and 2017 \cite{iyengar2018us}. A follow-up survey in 2022 also reported high poetry readership: 11.5\% of adults in the US engaged in poetry either by reading or listening, while 4.8\% of US adults engaged in poetry \textit{listening} \cite{ArtsGov2023}, highlighting a large audience for poetry audio. Recently, the Recording Academy introduced a new, dedicated category for ``Best Spoken Word Poetry Album" to the Grammy Awards  \cite{GrammyRoundtable2023}, as a response to calls from the broader spoken word community for more equitable representation of poetry reading. 
However, compared to the other oral performance types, e.g., narrations and singing voice, computational analysis of poetry reading's acoustical characteristics has rarely been done in the literature.  


Similarly to any other speech, poetry reading must also contain a certain level of musicality coming from the pitched voice and the pace of reading \cite{nooteboom1997prosody}. However, it also has a deep-rooted connection with vocal music or singing voice  \cite{francis2021verbal}. Historically, poetry has been conveyed through oral performances, or recitations, across various cultures. For instance, ancient Greek poets often performed lyric poems accompanied by a lyre; traditional Chinese poems were often sung to accompany a musical instrument, such as the pipa or the guqin \cite{harvey1955classification, hou2015analyzing}. Also, poetry often employs musical devices to enhance its rhythm. For example, rhyme creates correspondence in sounds between words, while meter dictates the rhythmic structure through patterns of stressed and unstressed syllables. Additionally, poetry reading exhibits pitch variations as a form of spoken language. However, it is not widely agreed upon how to interpret the pitched signal as a form of ``melody" or how its characteristics differ from those in singing voice or narrations \cite{scharinger2023melody}. 


Extensive research has been conducted on the acoustic features of narrated speech (e.g., spoken language that reads non-poetic text or conversational speech) and singing voice. For example, compared to narrating, singing tends to use higher pitch, slower temporal rate, and more stable pitches and rhythm \cite{vanden2023developmental, ozaki2022similarities, list1963boundaries, patel2006comparing}. However, few studies have attempted to quantify where poetry reading falls within this spectrum. Hence, there is a significant gap in our understanding while poetry reading appears to exhibit a unique blend of characteristics typical of both narrating and singing. Meanwhile, computational analyses of poetry in the literature have primarily been done from a natural language understanding perspective, e.g., understanding the semantics, themes, and meanings of the text \cite{kaplan2007computational, kao2012computational, rakshit2015automated, kaur2020designing, ChoiK2023poetry_theme, singhi2014poetry}, missing the acoustic aspect of poetry reading. 

In this paper, we conduct a large-scale quantitative analysis of three different types of oral performance: singing voice, narration, and poetry reading. To this end, we propose a few signal processing algorithms that allow us to scale up the experiments to the degree that was previously infeasible. Specifically, we begin with an analysis of silence patterns of the three categories, and then provide insights coming from their pitch variation patterns, e.g., whether the poetry reading has steady-pitched areas as in singing voice. Finally, we also provide an analysis of the beat stability of the three types and present our findings that the poetry reading has a certain level of established beat patterns. We perform the experiments on three large-scale, publicly-available datasets: poetry readings collected from The Poetry Foundation\footnote{https://www.poetryfoundation.org}, the LibriSpeech audiobook dataset \cite{PanayotovV2015LibriSpeech}, and the Intonation set with isolated singing voice recordings \cite{WagerS2019icassp1}. We ensured that other non-vocal sound sources, e.g., musical instruments, sound effects, etc., are not included in our datasets, while some recordings could contain moderate environmental noise. To our best knowledge, this paper reports the first large-scale quantitative analysis of poetry reading in comparison to narrations and singing voices. The proposed algorithms are developed to provide the best-effort analysis of the poetry reading, while we hope that deeper and broader quantitative studies follow our work for a better understanding of poetry reading. Whilst the study is at an unprecedented scale, it is still limited to Western pop music. Moreover, 93\% of the poems were written after 1980, confirming a strong contemporary focus. Since poems in this era use more flexible and experimental forms, e.g., free verse, \cite{baumann2018style}, the signal processing algorithms are more challenged to find a distinguishable feature. We open-source our project and provide necessary metadata needed to reproduce the results: \url{https://github.com/kc82/poetry-reading}.




\section{Methodology}
In the literature, both narration and singing voice have been acoustically examined through a variety of features, including pitch range, pitch stability, rhythm, and tempo \cite{ vanden2023developmental, ozaki2022similarities, list1963boundaries, patel2006comparing}. However, these studies were on small datasets and subjective measures such as user responses or author annotations for assessment. To scale up, we devise quantifiable metrics and apply them to more than 1,000 audio files for each category. We aim to position poetry reading within the spectrum that spans narration and vocal music. Although we analyze audio from a local timeframe to a global histogram perspective, we leave the consideration of long-term patterns to future work. 

\subsection{Preprocessing via Transcription}\label{sec:prepro}
We preprocess the audio signals using WhisperX to distinguish between voiced and silent segments, detect language, and select clean audio files \cite{bain2023whisperx}. Whisper \cite{radford2023robust} is a state-of-the-art automatic speech recognition (ASR) package renowned for its low word error rate (WER), whose successor WhisperX provides improved performance. We opt for WhisperX because it offers more accurate word boundary detection through its phoneme-based ASR, which provides the beginning and ending timestamps of each word. Additionally, it is equipped with voice activity detection (VAD), making it useful for identifying silent intervals within audio files. 

\noindent\textbf{Silence Detection Based on Word Boundaries}: Audio files often contain silence either at the beginning, end, or interspersed between voiced segments. These patterns of silence can vary across different audio genres. To consider or mitigate the impact of silence on some of our analysis algorithms, we conduct studies on both silence-removed and contained versions of the audio. A simple-minded silence detection approach, e.g., a decision based on a threshold amplitude, is not a robust approach as some of the recordings contain different levels of noise. More appropriate methods are based on the VAD model trained from noisy speech signals, such as the one WhisperX provides. Instead of using the VAD output directly, we further process the signal to compute the word boundaries using WhisperX's ASR module. Hence, we define silence by the areas outside of the word boundaries. This is at the cost of excluding some spoken words that ASR fails to detect or filler sounds with no meanings, making the result after silence removal more conservative. 

\noindent\textbf{Language Detection}: Each language has its distinct acoustic characteristics. To control for this variable and align with the English LibriSpeech dataset, our analysis is limited to English content. This is done by WhisperX's language detection score, i.e., if it is greater than 0.93 for English. However, we plan to include multiple languages in future studies. 




\subsection{Local Pitch Variability}
A well-known qualitative feature of the singing voice is its steady pitch that remains within a short period of time, i.e., the concept of musical notes. On the contrary, in narration, the vowel sound often varies its pitch over time, creating an unstable pitch contour \cite{vanden2023developmental, ozaki2022similarities}. In this paper, we empirically capture this local pitch variation. 

First, we estimate the pitch value at every 64 ms-long audio frame by using the probabilistic YIN (pYIN) algorithm \cite{MauchM2014pyin}. First of all, we ignore the silent areas defined by the ASR model as described in Sec. \ref{sec:prepro}. Yet, for some unvoiced frames, pYIN can fail and result in ``not a number" (NaN). Hence, when we compute the local pitch statistics of 12 consecutive frames (96 ms), we make sure that (a) there are less than seven NaN values (b) at least five consecutive pitch values are found within the 12-frames window. Otherwise, we disregard that 12-frame chunk. These local statistics (i.e., standard deviation of those pitch values) will tell us whether pitch varies too much within a short period of time or sustains. We will perform statistical tests to verify that the poetry reading is more likely to contain sustained pitches than narration. 

\subsection{Beat Stability}
The rationale behind the beat stability feature is that the beat tracking effort (either made by human listeners or machines) should be mitigated if the audio signal contains a regular beat pattern. Otherwise, such as in narration, it takes more exceptions to identify candidate beat locations that are off from the perceived regularity. 

To this end, we employ a well-known dynamic programming (DP) algorithm to track down the beats faithfully \cite{ellis2007beat}, which is implemented as the \texttt{librosa.beat.beat\_track} function in \texttt{librosa} \cite{mcfee2015librosa}. Specifically, its objective function provides a systematic way to quantify the rhythmic irregularity, which we use to compare the three categories. The objective is defined over the onset signal $O(\cdot)$ as input, which records the abrupt changes of audio, and then finds the beat sequence $\{t_i\}$ that maximizes the score function:
\begin{equation}\label{eq:beat_track_obj}
    C(\{t_i\})=\sum_{i=1}^N O(t_i) + \alpha \sum_{i=2}^N F(t_i - t_{i-1}, \tau_p).    
\end{equation}
The objective function can be understood from two perspectives. Since perceived beats are based on sound events, if the onset function's values at the beat positions are large, the beats match the salient sound events' locations. The first summation in the objective function quantifies this: the larger the sum is, the better the match is. 

Meanwhile, the objective function also allows some beat locations that are slightly off from the regularity. The beat tracking algorithm first finds the global tempo from the onset signal, and then tries to keep the number of beats per minute (BPM) the same as the global tempo. Here, the second term regularizes the optimization by enforcing a particular beat interval, $\tau_p$, to all inter-beat time differences $t_i - t_{i-1}$ using a pre-defined penalty function: $F(\Delta t, \tau)=- \left( \log \frac{\Delta t}{\tau} \right)^2$. Hence, if it were not for the second term, the first term only quantifies the sum of the onset function values and can produce a trivial solution, e.g., beat locations match the loudest sound events regardless of the periodicity. 

$\alpha$ is an important hyperparameter, which controls the ``tightness" of the beat estimates compared to the target tempo. A small $\alpha$ will result in less regular beat patterns, as the DP optimizer relies only on the large onset values (the first term of eq. \eqref{eq:beat_track_obj}). On the other hand, a too-large $\alpha$ will estimate a periodic sequence of beat locations that are irrelevant to the perceived beats defined by the sound events. 

In this work, we quantify the beat stability by using the DP algorithm's performance depending on the different choices of $\alpha$: if a steady beat pattern is observed, the objective function $C(\{t_i\})$ will be affected less by a smaller choice of $\alpha$, and vice versa. Therefore, for a given recording, we compare the total beat tracking scores by changing $\alpha$ from 1,000 to 1 (the default value is 100). As for the onset detection, we use the spectral fluctuation feature \cite{MasriP1996phd}. We denote the optimal objective function value of the audio signal by 
\begin{equation}\label{eq:max_DP_score}
    C^*=\argmax_{\{t_i\}}C(\{t_i\}).
\end{equation}

\section{Datasets}
\label{sec:datasets}

Table \ref{tab:datast} summarizes the basic information of the datasets we use in this paper. The datasets are closely matched in terms of both the number of files and word counts. 

\noindent\textbf{Poetry Reading Dataset}: The poetry reading dataset is sourced from \url{www.poetryfoundation.org}. The Poetry Foundation stands as a distinguished platform boasting thousands of audio recordings, offering a broad spectrum of data as well as the reliability of the dataset when analyzing the vocal nuances of poetry. We only consider the poems that are accompanied by audio, i.e., a recorded poetry recitation. Starting from the initial 1,300 crawled audio files, we refine the dataset by choosing the ones that WhisperX detects their language as English with 93\% certainty. Although it does not necessarily mean that those excluded samples were written in other languages, in this way, the selected examples are favored by the ASR model, improving the word boundary detection performance. This selection step reduces the final dataset down to 1,058 audio files. We provide the URLs of the poems we used in this study. We reduce their sample rate to 16kHz.



\noindent\textbf{Narration Dataset}: LibriSpeech \cite{PanayotovV2015LibriSpeech} provides a collection of 1,101 unique books translated into audio files with 16 kHz sample rate, serving as a representation of narrative speech. This is primarily because the content is drawn from full-length books, which inherently encompass continuous and extended narratives, both in fiction and non-fiction genres. The natural flow of language, varying tones, cadences, and expressive modulations intrinsic to book readings capture the essence of narrative speech. Furthermore, as these recordings are derived from diverse authors, styles, and periods, they collectively offer a comprehensive portrayal of storytelling and narrative techniques across a broad spectrum. We carefully concatenated the segmented audio signals to be slightly over 90 seconds to increase the length of the signals in our experiments, while preserving their sequence and continuity of the narrative. We brought data only from the clean data fold, leaving out non-clean data. 


\noindent\textbf{Singing Voice Dataset}: We use the Intonation dataset \cite{WagerS2019icassp1} as a representative of singing voice. Intonation contains 4,702 audio files sourced globally using a karaoke app serviced by Smule, Inc. This ensures a broad spectrum of singing styles, techniques, and nuances inherent to diverse cultures and traditions. The user-generated nature of Intonation brings a raw authenticity to the collection, encompassing both trained voices and natural, untrained vocal expressions. In addition, Intonation is suitable for our purposes because the recordings do not contain the other accompanying musical instruments. Once again, we reduce it down to 1,050 English-language songs using WhisperX's language detection with the same criteria applied to select poetry audio. We resample the signals at a 16kHz rate.



\section{Experiments}
\label{sec:experiment}

\subsection{Silence Patterns}
In all three categories, it is common to observe silent regions in between words. However, their lengths and functions in the oral performance have different meanings. For example, in narration, there tends to be a pause between sentences. On the other hand, in singing voice, long pauses are also common in Western pop music, e.g., during the interlude. We are interested in the intentional pauses in poetry reading, which are often expected after a line or stanza \cite{dillon1976clause} to maintain the rhythms or make poems more song-like \cite{lissa1964aesthetic}.

\begin{figure}[t]
     \centering
     \begin{subfigure}[b]{0.332\columnwidth}
         \centering
         \includegraphics[width=\textwidth]{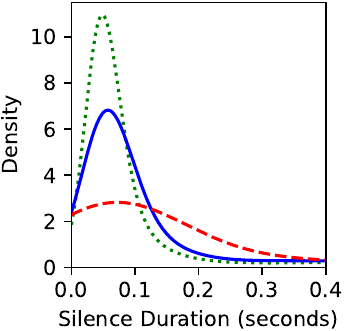}
         \caption{}
         \label{fig:short_silence}
     \end{subfigure}
     \hfill
     \begin{subfigure}[b]{0.311\columnwidth}
         \centering
         \includegraphics[width=\textwidth]{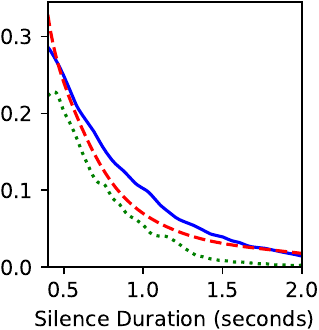}
         \caption{}
         \label{fig:medium_silence}
     \end{subfigure}
     \hfill
     \begin{subfigure}[b]{0.336\columnwidth}
         \centering
         \includegraphics[width=\textwidth]{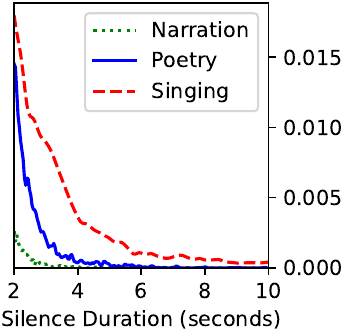}
         \caption{}
         \label{fig:long_silence}
     \end{subfigure}
     \vspace{-0.25in}
        \caption{Histograms of the (a) short (b) medium (c) long silent segments.}
\vspace{-0.2in}        
        \label{fig:silence}
\end{figure}

To capture this, we draw histograms of the lengths of the silent periods, which are acquired from the silence removal process in Sec. \ref{sec:prepro} as a byproduct. Fig. \ref{fig:short_silence}, shows three histograms of relatively short silent chunks (from 0 to 0.4 second). First, very short pauses (below 0.1 second) are the most common type across the three categories; however, their prevalence varies, being most frequent in narration, moderately common in poetry reading, and comparatively less frequent in singing voice. In Fig. \ref{fig:medium_silence}, we see more pauses in poetry reading, indicating that poetry reading contains those medium-length (around 1 second) pauses, more than other categories. Finally, In Fig. \ref{fig:long_silence}, the density of singing voice surges, showcasing its frequent long pauses. Likewise, we can see that moderate-length pauses are relatively common in poetry reading, which is a unique feature that differentiates it from other categories. 



\begin{table}[t]
    \centering
    \caption{Summary of the Datasets}
    \vspace{-0.1in}
    \resizebox{.9\columnwidth}{!}{%
    \begin{tabular}{lrrr}
    \toprule
    & LibriSpeech & Poetry Reading & Intonation \\
    \midrule
    Number of Files & 1,101 & 1,058 & 1,050 \\
    Spoken Duration (min) & 1,079 & 1,318 & 1,889 \\
    Word Count & 285,233 & 307,039 & 269,567 \\
    Total Silence (min) & 653 & 1,142 & 1,450 \\
    Words Per Min. & 164.73 & 124.88 & 80.76 \\
    Std. Silence (sec) & 0.25 & 0.41 & 1.25 \\
    \bottomrule
    \end{tabular}%
    }
    \label{tab:datast}
    \vspace{-0.18in}
\end{table}

\begin{figure}[t]
    \begin{subfigure}[b]{\columnwidth}
         \centering
         \includegraphics[width=\textwidth]{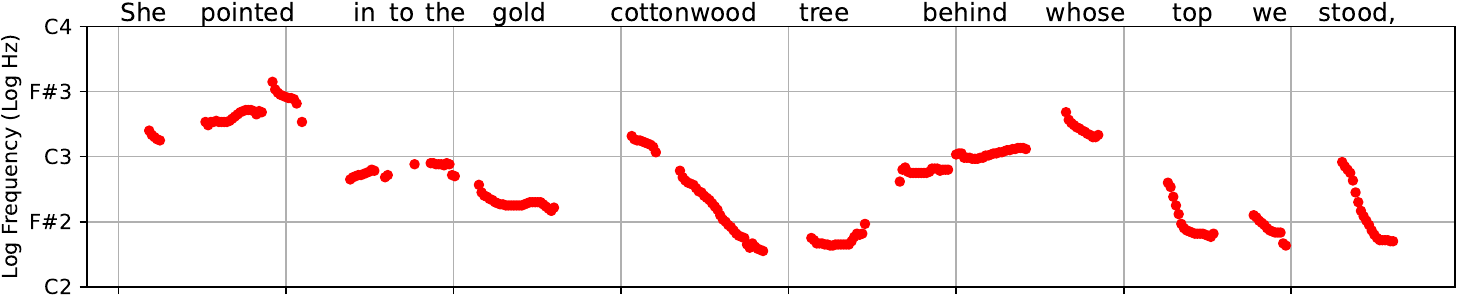}
         \includegraphics[width=\textwidth]{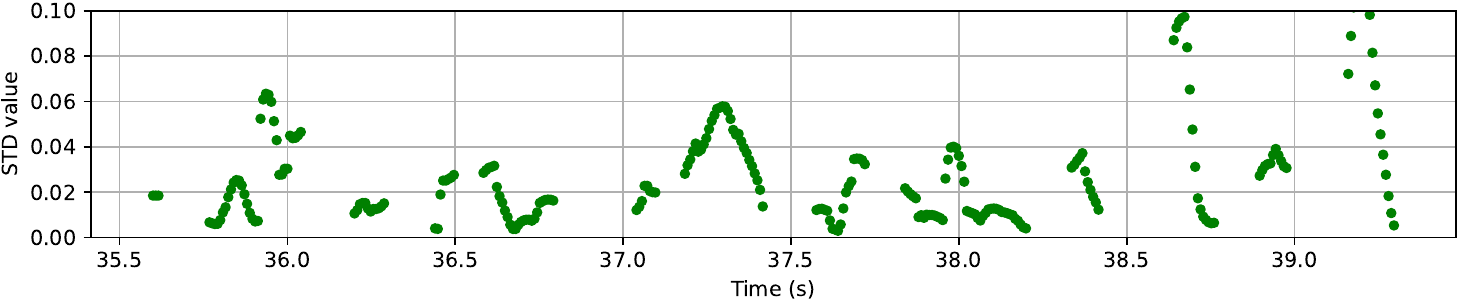}
         \vspace{-0.2in}
         \caption{Narration}
         \label{fig:pitch-narration}
     \end{subfigure}
    \begin{subfigure}[b]{\columnwidth}
         \centering
         \includegraphics[width=\textwidth]{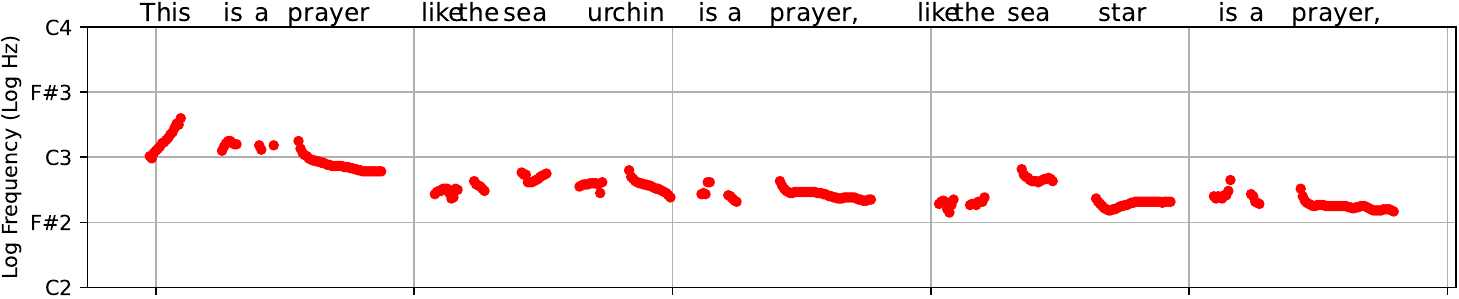}
         \includegraphics[width=\textwidth]{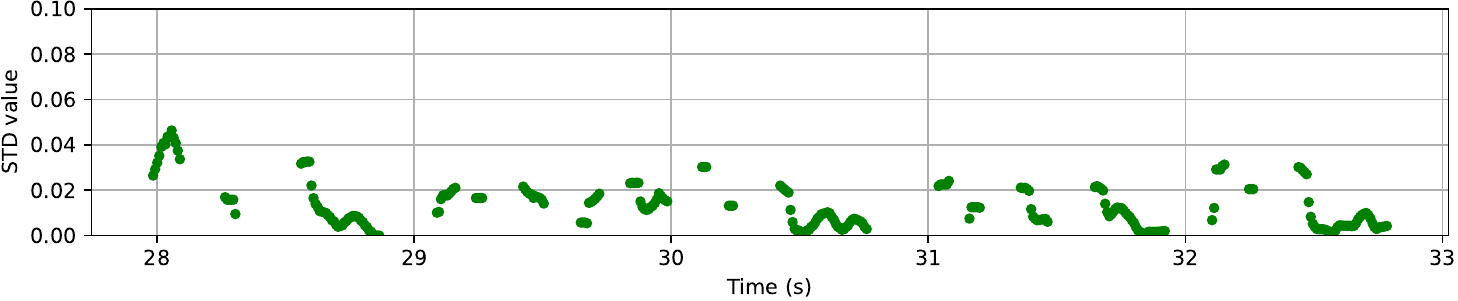}
         \vspace{-0.2in}
         \caption{Poetry reading}
         \label{fig:pitch-poetry}
     \end{subfigure} 
    \begin{subfigure}[b]{\columnwidth}
         \centering
         \includegraphics[width=\textwidth]{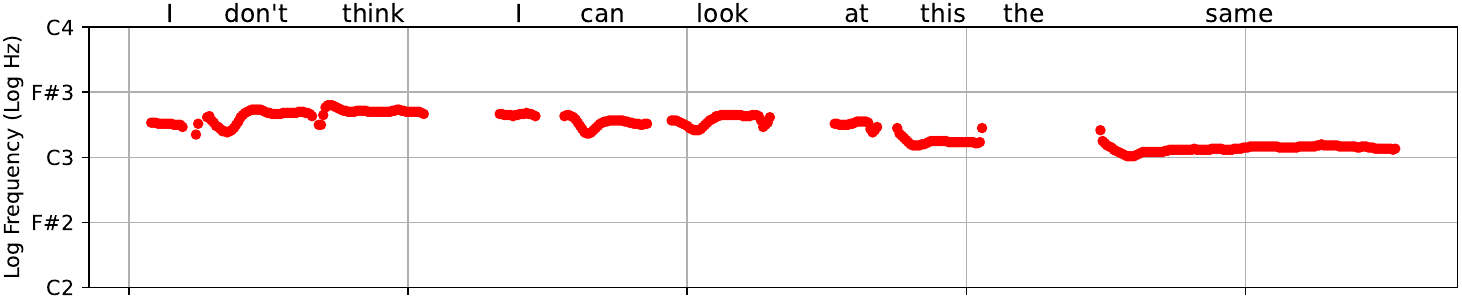}
         \includegraphics[width=\textwidth]{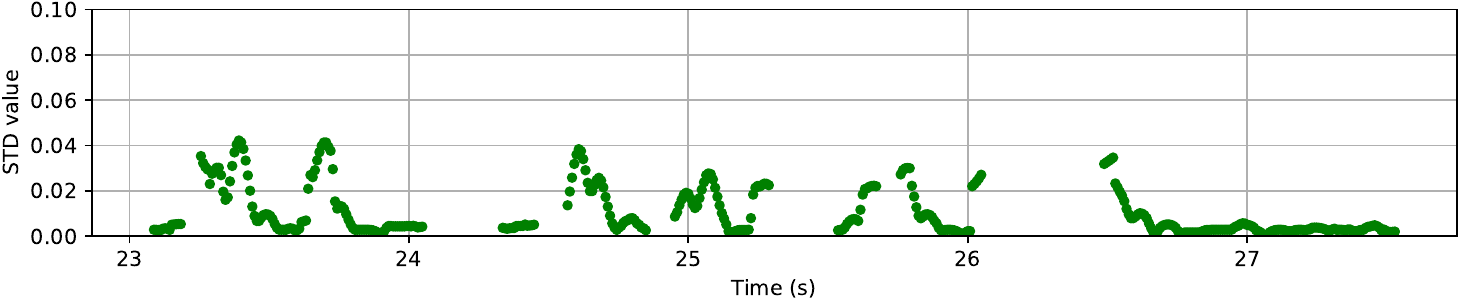}
         \vspace{-0.2in}
         \caption{Singing voice}
         \label{fig:pitch-vocal}
     \end{subfigure}  
     \vspace{-0.2in}
        \caption{Pitch contours (top) and local standard deviation (bottom).}
        \label{fig:pitch}
\end{figure}

\begin{figure}[t]
    \centering
    \hspace{0.1in}
    \includegraphics[width=0.93\columnwidth]{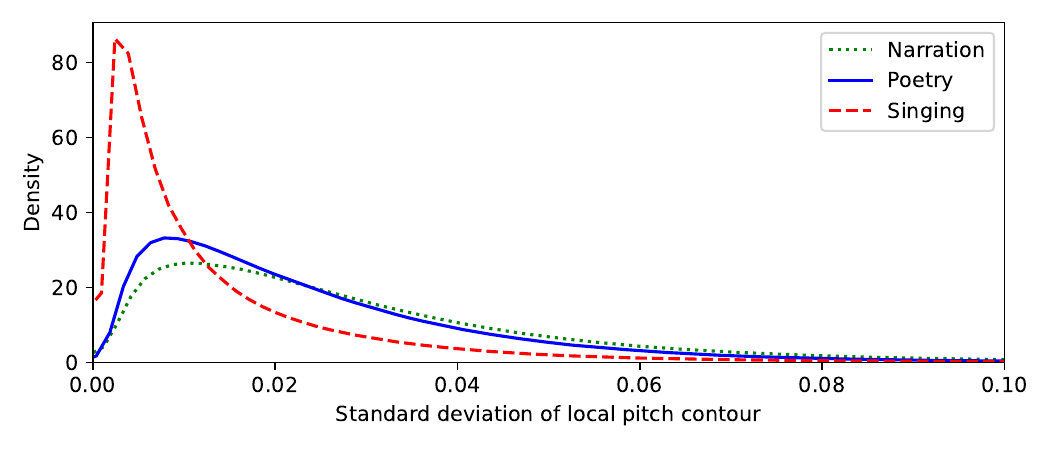}
    \vspace{-0.1in}
    \caption{Histograms of the std values of the local pitch contours.}
    \label{fig:pitch-hist}
\end{figure}

\subsection{Local Pitch Variability}



The top figures in Fig. \ref{fig:pitch} illustrate five second-long representative pitch contours (in red) of the three categories. Once again, in general, poetry reading and narration share similar patterns, i.e., less steady pitch contours compared to singing voices'. However, poetry reading sometimes contains more elongated vowel sounds and more steady-pitch areas than narration. The bottom graphs (in green) show the corresponding local pitch variation values, i.e., std of the 12 consecutive pitch values, where the high std values are correlated with the steep changes of the pitch contours (in narration), while the low-std regions are associated with the sustained musical notes (in singing voice). Poetry reading shows less drastic changes in its contour, leading to lower std than narration.

The histograms of the three std categories are shown in Fig. \ref{fig:pitch-hist}. Once again, poetry reading shows a similar trend to narration, while it has more density towards zero. We conduct a two-sample Kolmogorov-Smirnov (KS) test \cite{KS-test} to verify the moderate but statistically significant difference between poetry reading and narration. Their KS statistic of 0.089 is relatively small, while the p-value is 0.0 (below the machine precision): the chance of the two std sets being from the same distribution is improbable.

\begin{figure}[t]
     \centering
     \begin{subfigure}[b]{0.99\columnwidth}
         \centering
         \hspace{0.04in}
         \includegraphics[width=0.945\textwidth]{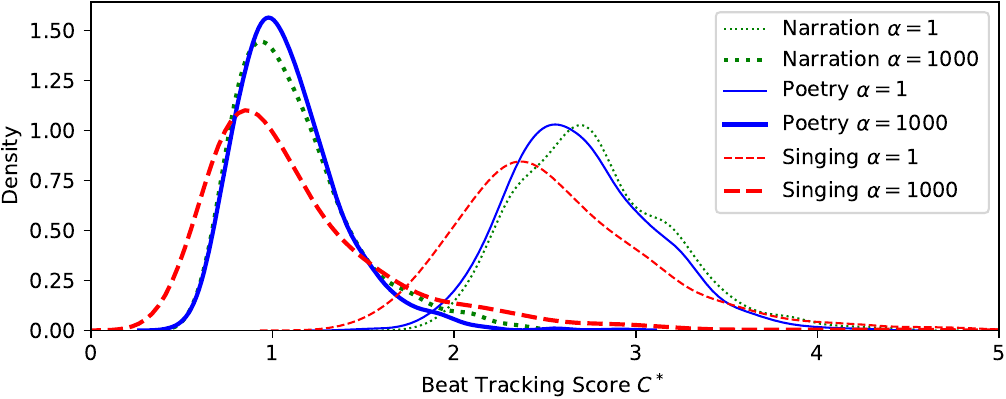}
         \vspace{-0.05in}
         \caption{Change of the beat tracking score $C^*$ by varying the tightness value $\alpha$}
         \label{fig:beat_track_all}
     \end{subfigure}
     \hfill
     \vspace{0.1in}
     \begin{subfigure}[b]{0.99\columnwidth}
         \centering
         \hspace{0.08in}
         \includegraphics[width=0.94\textwidth]{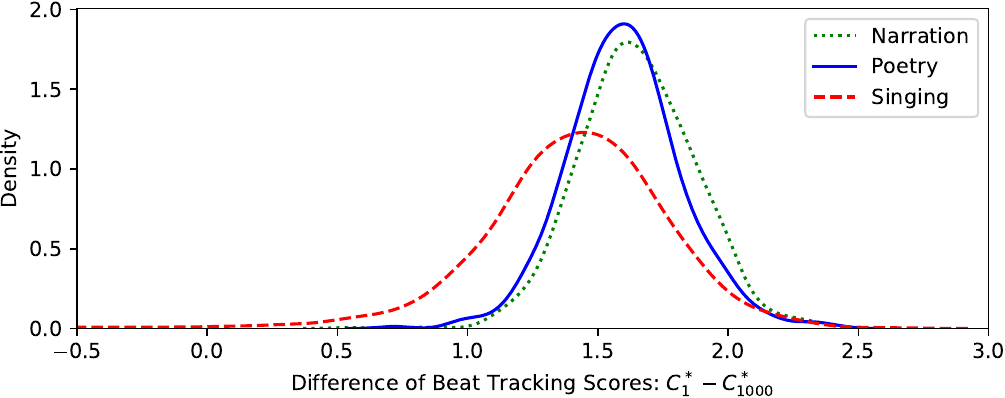}
         \vspace{-0.05in}
         \caption{Histogram of the difference of the beat tracking scores $C^*_1-C^*_{1000}$}
         \label{fig:beat_track_diff}
     \end{subfigure}
        \caption{Histograms of the beat tracking scores.}
        \label{fig:beat_track}
\end{figure}

\subsection{Beat Stability}

We extract the final score $C^*$ as defined in eq. \eqref{eq:max_DP_score}, which indicates the beat-tracking algorithm's fitness for the signal. We consider two potential issues to interpret the scores. First, $C^*$ keeps increasing as the algorithm finds more beats from the signal, so we divide the final score by the number of found beats to normalize it. 

Second, due to the variations among different recordings, a direct comparison of the scores computed from two different examples may not be robust. For example, a soft song with a regular beat pattern could have a lower score than a series of drum attacks with no periodicity due to their differences in the onset peaks. To this end, we propose to compute two beat-tracking scores per recording by using two different tightness setups, $\alpha=1$ and $1,000$, i.e., $C^*_1$ and $C^*_{1000}$. Our assumption is that, even for the same audio, $C^*_1$ will go up because beat tracking is free to find the largest onset peaks, ignoring the inter-beat temporal regularity. However, the same audio's $C^*_{1000}$ could be poor as there is less chance for the beat tracking algorithm to find the onset peaks at the rigidly regular beat intervals, unless the audio signal already contains a regular beat pattern.

Fig. \ref{fig:beat_track_all} presents two histograms from $C^*_1$ (thin line) and $C^*_{1000}$ (thick line) per category. In all three categories, we see that $C^*_{1}$'s distribution is overall with higher scores than $C^*_{1000}$ as expected. However, their differences vary depending on the category. For example, the difference between $C^*_{1000}$ and $C^*_{1}$'s empirical distributions for the singing voice category is the smallest, with a Wasserstein distance \cite{Wasserstein} of 1.41. In other words, the singing voice tends to contain more regular beat patterns, which can be found more easily by the beat tracking algorithm, even with the very tight regularization. On the other hand, narration's distributions change more drastically with a higher Wasserstein distance of 1.65: the algorithm exploits less regularization to find higher onset peaks. What is interesting is the poetry reading distributions: their beat tracking results differ more widely than singing voice's by varying $\alpha$, but their difference is less than that of narration (1.60). It means the very rigid enforcement of a steady tempo still finds more meaningful beat locations from poetry reading than narration. 

In Fig. \ref{fig:beat_track_diff}, we compute the difference between $C^*_{1}$ and $C^*_{1000}$ of each recording, and draw their histograms. It re-emphasizes our findings (a) singing voice is less affected by the beat tracking algorithm's rigidness than narration (b) poetry reading contains slightly, but significantly more beat patterns. The KS test between the narration and poetry distributions from this graph reports a relatively low score of 0.1059, reflecting their resemblance. However, the p-value is significantly low ($9.896 \times 10^{-6}$), thus clearly rejecting the hypothesis that they originate from the same distribution.

\section{Conclusion}
In this research, we applied computational methodologies to analyze the acoustic parameters of professionally-read poetry, a domain previously less scrutinized than narrative speech and singing voice. Utilizing advanced signal processing techniques, we specifically addressed the characteristics of poetry reading based on silence patterns, pitch temporal variations, and beat stability. The analysis results from three comprehensive corpora indicated that poetry reading exhibits intermediate characteristics between narrative and singing. To our best knowledge, this is the first large-scale poetry reading analysis. We open-sourced our project for successive research.

\bibliographystyle{IEEEbib}
\bibliography{strings}

\end{document}